\documentclass[10pt,conference,a4paper,twocolumn]{IEEEtran}
\IEEEoverridecommandlockouts

\IEEEpubid{
    \makebox[\columnwidth]{979-8-3315-8229-6/26/\$31.00~\copyright{}2026 IEEE \hfill}
    \hspace{\columnsep}\makebox[\columnwidth]{ }
}

\usepackage{cite}
\usepackage{amsmath,amssymb,amsfonts}
\usepackage{algpseudocodex, algorithm}
\usepackage{graphicx}
\usepackage{textcomp}
\usepackage{pbalance}
\usepackage{siunitx}
\usepackage[switch]{lineno}
\usepackage[sort=none]{glossaries}
\usepackage{csquotes}
\usepackage{paralist}
\usepackage{lipsum}
\usepackage{multirow}
\usepackage{booktabs}
\usepackage{tabularx}
\usepackage{fontawesome5}

\usepackage[absolute,showboxes]{textpos}

\usepackage{todonotes}
\usepackage{caption}
\usepackage[position=b]{subcaption}

\usepackage[hidelinks]{hyperref}

\newif\ifanonymous

\usepackage{titlesec}
\titlespacing*{\section}{0pt}{6pt}{4pt}
\titlespacing*{\subsection}{0pt}{4pt}{3pt}

\let\orgautoref\autoref
\renewcommand{\autoref}
{\def\sectionautorefname{Section}\def\subsectionautorefname{Section}\def\subsubsectionautorefname{Section}\def\figureautorefname{Fig.}\def\equationautorefname{Eq.}\def\algorithmautorefname{Algorithm}\orgautoref}

\usepackage{pifont}

\usepackage{xspace}
\newcommand{\etal}{\textit{et al.}~}
\newcommand{\eg}{\textit{e.g.,}~}

\newcommand{\cf}{\textit{cf.}~}

\makeatletter
\renewcommand{\paragraph}[1]{\vspace*{0.03in}\noindent{\bf #1.}\hspace{0.25ex \@plus1ex \@minus.2ex}}
\makeatother

\makeatletter
\newcommand{\paragraphc}[1]{\vspace*{0.03in}\noindent{\bf #1}\hspace{1ex \@minus.2ex}}
\makeatother

\newcommand{\numofdevs}{eight\xspace}

\newcommand{\zerostars}{{\faStar[regular] \faStar[regular] \faStar[regular]}}
\newcommand{\onestar}{{\faStar \xspace \faStar[regular] \faStar[regular]}}
\newcommand{\onehalfstars}{{\faStar \xspace \faStarHalf* \faStar[regular]}}
\newcommand{\twostars}{{\faStar \xspace \faStar \xspace \faStar[regular]}}
\newcommand{\twohalfstars}{{\faStar \xspace \faStar \xspace \faStarHalf*}}
\newcommand{\threestars}{{\faStar \xspace \faStar \xspace \faStar}}

\IEEEoverridecommandlockouts

\setacronymstyle{long-short}
\newacronym{age-check}{AGE-WATCH}{Ageing Window-based Analysis and Timing CHeck}
\newacronym{bti}{BTI}{bias temperature instability}
\newacronym{cots}{COTS}{commercial off-the-shelf}
\newacronym[longplural=devices under test]{dut}{DUT}{device under test}{}
\newacronym{fet}{FET}{field effect transistor}
\newacronym{fpga}{FPGA}{field programmable gate array}
\newacronym{hci}{HCI}{hot carrier injection}
\newacronym{ic}{IC}{integrated circuit}
\newacronym{iqr}{IQR}{interquartile range}
\newacronym{mef}{MEF}{maximum error-free frequency}
\newacronym{mof}{MOF}{maximum operational frequency}
\newacronym{mcu}{MCU}{microcontroller unit}
\newacronym{sbst}{SBST}{software-based self-testing}
\newacronym{sram}{SRAM}{static random access memory}
\newacronym{tddb}{TDDB}{time-dependent dielectric breakdown}

\makeatletter \newcommand{\linebreakand}{\end{@IEEEauthorhalign}
  \hfill\mbox{}\par
  \mbox{}\hfill\begin{@IEEEauthorhalign}
}
\makeatother

\begin{document}

\bstctlcite{IEEEexample:BSTcontrol}

\setlength{\TPHorizModule}{\paperwidth}
\setlength{\TPVertModule}{\paperheight}
\TPMargin{5pt}
\begin{textblock}{0.8}(0.1,0.02)
     \noindent
     \footnotesize
     If you cite this paper, please use the DDECS reference:
     Leandro Lanzieri, Jiri Kral, Goerschwin Fey, Holger Schlarb, and Thomas C. Schmidt.
     Ageing Monitoring for Commercial Microcontrollers Based on Timing Windows.
     In \emph{Proceedings of the 29th IEEE International Symposium on Design and Diagnostics of Electronic Circuits and Systems (DDECS)}, IEEE, 2026.
\end{textblock}

\title{
    Ageing Monitoring for Commercial Microcontrollers Based on Timing Windows
    \ifanonymous
    \else
    {\footnotesize
        \thanks{
            We acknowledge the support by DASHH (Data Science in Hamburg — Helmholtz Graduate School for the Structure of Matter) with the Grant-No. HIDSS-0002,
            the Federal Ministry of Education and Research with Grant C-ray4edge No. 16KIS1695, and the DFG project No. 47183717.
        }
    }
    \fi
}

\ifanonymous
\author{\IEEEauthorblockN{Paper \#NNN, \pageref{lastpage}~pages (with references)}}
\else

\author{\IEEEauthorblockN{Leandro Lanzieri}
    \IEEEauthorblockA{\textit{Deutsches Elektronen-Synchrotron DESY} \\
        Hamburg, Germany \\
        leandro.lanzieri@desy.de
    }
    \and
    \IEEEauthorblockN{Jiri Kral}
    \IEEEauthorblockA{\textit{Deutsches Elektronen-Synchrotron DESY} \\
        Hamburg, Germany \\
        jiri.kral@desy.de
    }
    \and
    \IEEEauthorblockN{Goerschwin Fey}
    \IEEEauthorblockA{\textit{Hamburg University of Technology} \\
        Hamburg, Germany \\
        goerschwin.fey@tuhh.de
    }
    \and

    \linebreakand

    \IEEEauthorblockN{Holger Schlarb}
    \IEEEauthorblockA{\textit{Deutsches Elektronen-Synchrotron DESY} \\
        Hamburg, Germany \\
        holger.schlarb@desy.de
    }
    \and
    \IEEEauthorblockN{Thomas C. Schmidt}
    \IEEEauthorblockA{\textit{Hamburg University of Applied Sciences} \\
        Hamburg, Germany \\
        t.schmidt@haw-hamburg.de
    }
}
\fi

\maketitle

\vspace*{-1cm}

\begin{abstract}
    Microcontrollers are increasingly present in embedded deployments and dependable systems, for which malfunctions due to hardware ageing can have severe impact.
    The lack of deployable techniques for ageing monitoring on these devices has spread the application of guard bands to prevent timing errors due to degradation.
    Applying this static technique can limit performance and lead to sudden failures as devices age.
    In this paper, we follow a software-based self-testing approach to design monitoring of hardware degradation for microcontrollers.
    Deployable in the field, our technique leverages timing windows of variable lengths to determine the maximum operational frequency of the devices.
    We empirically validate the method on real hardware and find  that it consistently detects temperature-induced degradations in maximum operating frequency of up to 13.79\% across devices for 60 °C temperature increase.
\end{abstract}


\section{Introduction}

\Gls{cots} \glspl{mcu} are ubiquitous in dependable embedded applications thanks to their versatility, affordability, and low energy usage.
During operation, \glspl{mcu} undergo hardware degradation owing to ageing mechanisms that impact underlying transistors, increasing the propagation delay of signals, reducing operating frequencies, and potentially causing timing errors \cite{2024_Lanzieri_StudyingTheDegradation,2023_Lanzieri_AgeingAnalysisOf}.
\Gls{mcu} vendors impose operational guard bands on clock frequencies to account for ageing by considering worst-case conditions, which can limit performance and lead to sudden failures \cite{2022_Kaushik_EvaluationOfDynamic}.
Techniques that assess ageing at runtime for \gls{cots} \glspl{mcu} without circuitry changes lack exploration \cite{2024_Lanzieri_AReviewOf}, limiting the detection of ageing on \glspl{mcu} to approaches requiring chip modifications or external equipment \cite{2019_Hong_AVariationResilient,2024_Kaneko_ExperimentalEvaluationFor,2023_Safa_CounterfeitChipDetection}.

System-level tests are used during design, in particular \gls{sbst}, which allows processors to test themselves with programs \cite{2010_Psarakis_MicroprocessorSoftwareBased}.
Functional \gls{sbst} is convenient for \gls{cots} devices as it requires neither gate-level knowledge nor special cores, and can be deployed in the field.
System-level tests are applied to \gls{mcu} fault detection in embedded systems \cite{2007_Tamandl_OnlineSelfTests,2005_Paschalis_EffectiveSoftwareBased}, and research shows potential for automatic generation \cite{2003_Corno_FullyAutomaticTest}.

\begin{figure}[t]
    \centering
    \includegraphics[width=0.95\linewidth]{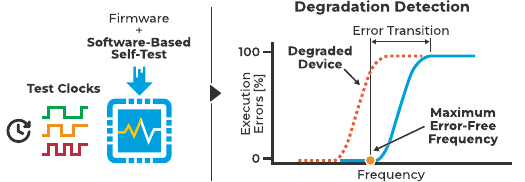}
    \caption{By observing the behaviour of payload execution at different frequencies, hardware degradation can be detected.}
    \label{fig:frequency_exploration}
    \vspace*{-0.5cm}
\end{figure}

In this paper, we propose an \gls{sbst} technique through which commercial \glspl{mcu} self-assess and monitor hardware ageing, as illustrated in \autoref{fig:frequency_exploration}.
By infrequently executing a self-test payload at varying clock frequencies, devices can estimate and track degradation in maximum operational frequency over time.
The periodic determination of this limit provides devices with a degradation indicator and enables ageing-aware decisions, such as acceptable overclocking tolerance.

We empirically evaluate our method by measuring changes of physical limits using \numofdevs commercial \glspl{mcu} and various firmware payloads at increasing temperatures, which we show to be an ageing proxy.
Results demonstrate that the technique successfully detects hardware degradation, which can be directly exploited by users deploying \gls{cots} devices.
Moreover, we complement the technique with a framework to compare computing payloads executed during the test.
Although \gls{sbst} and functional tests are well established for \gls{cots} \glspl{mcu}, this study---to the best of our knowledge---is the first to use them for online ageing monitoring.
In detail, this work contributes:
\begin{inparaenum}[i)]
    \item a deployable technique to detect ageing on \glspl{mcu} (\autoref{sec:methodology}),
    \item an empirical evaluation of the method,
    \item and an analysis of exemplary payloads and the system-level effects of degradation on \glspl{mcu} (\autoref{sec:experiment}).
\end{inparaenum}
\IEEEpubidadjcol

 \vspace*{-0.05cm}
\section{Background and Related Work}\label{sec:background}

\subsection{Hardware Ageing on Digital Devices}

The degradation of electronic devices during operation via mechanisms such as \gls{bti} and \gls{hci} leads to increased threshold voltage, and reduced carrier mobility and drain current.
These transistor variations impact performance and reliability of the devices by reducing their maximum operating frequencies (\(f_{\text{max}}\)) \cite{2024_Lanzieri_AReviewOf}.

In CMOS circuits, the propagation time (\(t_{\text{p}} = 1 / f_{\text{max}}\)) of a gate driving a load is inversely proportional to the drain current of its transistors, which in saturation is given by
\begin{equation}
    I_{\text{D}} = \frac{1}{2} \, \mu \, C_{\text{ox}} \left(\frac{W}{L}\right) \left(V_{\text{DD}} - V_{\text{th}}\right)^2
    \label{eq:transistor_current}
\end{equation}
where the oxide capacitance (\(C_{\text{ox}}\)), the channel width (\(W\)) and length (\(L\)), and the power supply voltage (\(V_{\text{DD}}\)) are fixed by design, but the carrier mobility (\(\mu\)), and the threshold voltage (\(V_{\text{th}}\)) vary over time \cite{2007_Schroder_NegativeBiasTemperature}.
Therefore, a weaker \(I_{\text{D}}\) reduces \(f_\text{max}\), as charging and discharging the load capacitance takes longer.

\subsection{Timing Windows and Guard Bands}

\begin{figure}[t]
    \centering
    \includegraphics[width=0.9\linewidth]{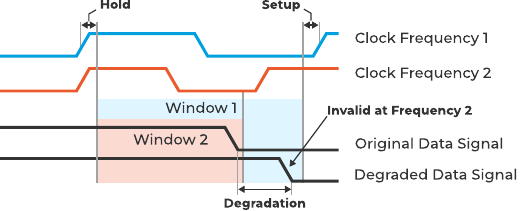}
    \caption{Higher frequencies give signal propagation shorter windows, triggering invalid transitions in case of degradation.}
    \label{fig:timing_window_variation}
    \vspace*{-0.4cm}
\end{figure}

Combinational logic in microcontrollers has to comply with timing requirements to ensure correct data propagation.
As illustrated in \autoref{fig:timing_window_variation}, two periods constrain the maximum allowed path delay: a setup time before the clock edge, at which data must be valid, and a hold time after the clock edge, during which the input signal must be stable.
The hold, setup, and clock periods define a \emph{timing window}, within which combinational circuits can transition to prevent timing errors.

As propagation delay of gates increases, the probability of invalid transitions grows.
To account for degradation, vendors define maximum allowed clock frequencies (guard bands) \cite{2019_Amrouch_OnTheEfficiency}, so that circuits comply with the windows even under variations.
Window lengths change with clock frequency: lower frequencies have larger windows and give signals more time to propagate, while higher frequencies shrink windows and tighten gate requirements.
Vendors statically define guard bands with tolerance for typical degradations under normal conditions during a guaranteed lifetime.
Since guard bands apply to all devices equally, and only coarsely account for manufacturing variations after binning, they can limit systems for which exploiting frequency margins improves performance.
Kaushik \etal evaluated dynamic frequency control on an automotive \gls{mcu}, combining overclocking for compute-intensive bursts with underclocking for less demanding tasks.
They achieved up to \qty{49}{\percent} energy savings thanks shorter execution times, faster transitions to idle, and less leakage \cite{2022_Kaushik_EvaluationOfDynamic}.

\subsection{Detecting Hardware Degradation on Microcontrollers}

Despite their wide deployment, \gls{cots} \glspl{mcu} lack research on deployable ageing detection \cite{2024_Lanzieri_AReviewOf}.
Due to non-disclosed implementations, detection techniques typically rely on side-channel measurements and functional testing.
M{\"u}hlbauer \etal proposed an \gls{sbst} technique based on error correction codes to manage the effects of flash ageing through graceful degradation rather than direct monitoring \cite{2017_Muhlbauer_HandlingManufacturingAnd}.
Akah \etal evaluated the resilience of \glspl{mcu} to total ionizing dose \cite{2017_Akah_TotalIonizingDose} by monitoring the frequency of pulses generated by a test application, but found the output stabilized beyond a certain dose, showing the method limitations.
Diggins \etal also studied the impact of radiation on \glspl{mcu}, focusing on timing violations \cite{2014_Diggins_TotalIonizingDose}.
The authors evaluated the maximum operational frequency of the devices at each dose, and showed that radiation degraded the propagation delay due to variations in threshold voltage and leakage currents.
Past a certain radiation level, devices could no longer operate at their nominal frequency and required slower clocks, clearly showing one of the limitations of static guard bands on dynamic systems.

Ageing estimation has also been applied to detect counterfeits: Safa \etal utilized the power delivery network of artificially aged \glspl{mcu} \cite{2023_Safa_CounterfeitChipDetection}, finding that variations in transistor capacitance and transconductance significantly impacted the network impedance, measurable via a vector network analyser.
Kaneko \etal observed emission changes after subjecting \glspl{mcu} to increased temperature and voltage \cite{2024_Kaneko_ExperimentalEvaluationFor}.

Most of the techniques for ageing monitoring on \gls{cots} \glspl{mcu} require complex measurement instruments, making them not deployment-friendly.
In this work, we follow the approach of \gls{sbst} for system-level tests on \glspl{mcu} \cite{2007_Tamandl_OnlineSelfTests}, and propose a method for self-assessment of hardware ageing.
We utilize timing window violations as a degradation indicator, which can be measured in the field.
Ageing-aware systems can use this insight for dynamic adaptations, thus optimizing performance or energy consumption while ensuring reliability.

\begin{figure*}[t]
    \centering
    \includegraphics[width=0.95\linewidth]{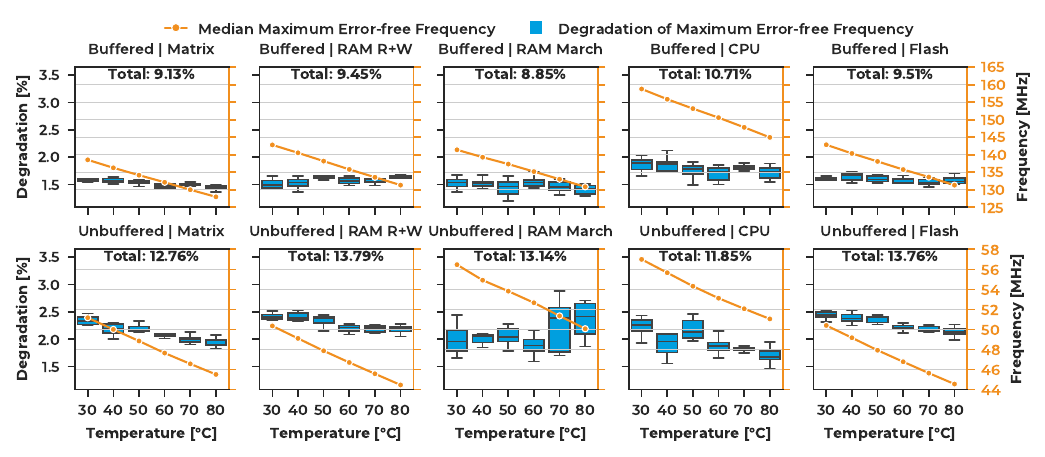}
    \vspace*{-0.3cm}
    \caption{Maximum error-free frequency and its degradation with respect to the previous temperature step. Degradation values are normalized to the maximum error-free frequency at \qty{20}{\degreeCelsius}. Columns correspond to payloads, and rows to buffered and unbuffered configurations. The total degradation at the top of each plot is the median value across devices.}
    \label{fig:degradation_err_free_freq_vs_temp}
    \vspace*{-0.3cm}
\end{figure*}

\section{Estimation of Microcontroller Degradation}\label{sec:methodology}

We propose estimating the degradation of propagation delay on commercial \glspl{mcu} by studying the execution of a known firmware payload at increasing clock frequencies.
The idea is to operate the devices on a controlled clock during test, thus to manipulate the length of timing windows (\cf \autoref{fig:timing_window_variation}).
Reducing the window length with a higher clock frequency ignores guard bands and triggers errors in the tests due to timing violations.
Depending on the degradation, payloads fail at different frequencies: more severe ageing induces errors at lower frequencies, due to longer propagation delays (\cf \autoref{fig:frequency_exploration}).
By periodically testing maximum frequencies, devices can indirectly monitor degradation.
The method requires typical elements of \gls{mcu} deployments: a controllable clock, persistence, and a watchdog or similar to detect timeouts.

\subsection{Exploring the Space of Clock Frequencies}

The test controller---typically the \gls{mcu} itself---generates the required clock and triggers the self-test.
The \gls{dut} uses the operation clock on standby, and switches to the test clock before the \gls{sbst} execution.
After running a payload, the \gls{mcu} returns to the standby configuration to verify the result and update an error count.
This allows the device to independently evaluate the test, which is required for a field deployment.
In this work, we show that a binary error result suffices for degradation detection, but further diagnostics are possible by analysing failures in detail.

The test controller explores a frequency space to profile the error behaviour of the payload.
As illustrated in \autoref{fig:frequency_exploration}, the main focus is the \gls{mef}: highest frequency at which the payload execution has no errors.
The \gls{mef} search follows an iterative bisection, in which the controller halves the space between a minimum frequency \(f_{\text{min}}\) and a maximum frequency \(f_{\text{max}}\), up to a step size \(\Delta f_{\text{b}}\), which is faster than a simple swipe.
Successful test runs conclude with an error report.
In case of a failed run, a timeout power-cycles the \gls{dut} (\eg via a watchdog).

\subsection{Evaluated Payload Probes}

An \gls{mcu} checks each frequency for errors by executing a payload multiple times.
Since the activated critical paths depend on the payload, the specific hardware and application determine the utilized payload.
\autoref{sec:experiment} describes metrics for users to compare and choose payloads.
In this study, we select a set of exemplary payloads that make heavier utilization of certain circuits, namely: matrix multiplication, flash read, RAM read and write (R+W), RAM march-C, and CPU test.

The CPU and RAM March-C tests are implementations of the class-B standard \cite{2022_Comission_IEC607301}.
The CPU test evaluates functionalities of the ALU (\eg flags) and registers.
RAM March-C is a \gls{sbst} that marches the memory in ascending and descending order writing and reading to uncover stuck-at, transition, address decoder, and coupling faults \cite{2001_Li_MarchBasedRAM}.
To reduce stress on the flash, these tests are executed from RAM.

The RAM R+W test heavily uses the memory by writing a known pattern, and verifying it with an MD5 hash.
Similarly, the flash test reads out known values, and verifies them.
Finally, the matrix test balances the usage of flash, RAM, and CPU by loading matrices into RAM, multiplying them, and calculating the determinant.
This test has been effective for error injection \cite{2009_Rohani_AnAnalysisOf} and undervolting \glspl{mcu} \cite{2016_Kulau_IdealvoltingReliableUndervolting}.  \section{Evaluation of Experiments}\label{sec:experiment}

\subsection{High Temperature as an Ageing Proxy}

Temperature negatively affects the performance of electronics by impacting carrier mobility in transistors \cite{2012_Wolpert_ManagingTemperatureEffects}.
At high temperatures, mobility is phonon-limited \cite{2007_Sze_PhysicsOfSemiconductor}: Phonons---quanta of vibrations in the lattice caused by thermal energy---scatter carriers as they move.
Since thermal energy increases vibration, there is an inverse power-law dependence between temperature and phonon-limited mobility \cite{1992_Klaassen_AUnifiedMobility}.

From \autoref{eq:transistor_current}, a temperature-induced reduction of mobility leads to a lower current.
Kumar \etal showed the impact on propagation delay in CMOS circuits, particularly on smaller node technologies \cite{2006_Kumar_ImpactOfTemperature}.
Ultimately, higher temperatures reduce the devices maximum frequency \cite{2003_Borkar_ParameterVariationsAnd, 2019_Amrouch_OnTheEfficiency}.

To experimentally validate our approach, we utilize \numofdevs commercial \glspl{mcu}: by applying temperature increments (\qtyrange{20}{80}{\degreeCelsius} in steps of \qty{10}{\degreeCelsius}), we leverage the temporary induced degradation \cite{2003_Borkar_ParameterVariationsAnd}.
This allows assessing the sensitivity of payloads and configurations to delay variations, using high temperature as a proxy for hardware ageing.

\begin{figure*}[t]
    \centering
    \includegraphics[width=0.9\linewidth]{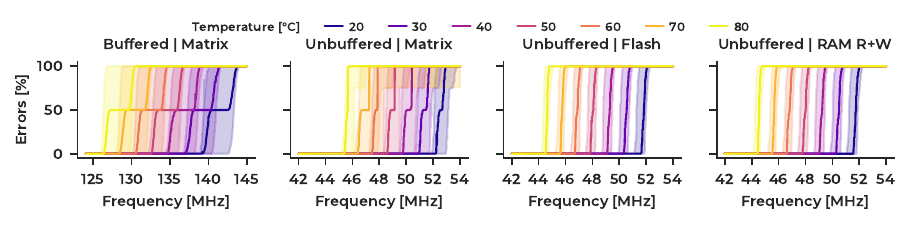}
    \vspace*{-0.3cm}
    \caption{Evolution of execution errors with frequency at different temperatures. Payloads without transition are omitted.}
    \label{fig:error_transition}
    \vspace*{-0.4cm}
\end{figure*}

\subsection{Experiment Setup}

\Glspl{dut} are 32-bit STM32F103RB \glspl{mcu} with ARM Cortex-M3 cores in \qty{130}{\nano\meter} technology, with a maximum system clock of \qty{72}{\mega\hertz}, and an external clock of up to \qty{25}{\mega\hertz}.
For test frequencies, we employ an SRS CG635 clock generator with a stability better than \qty{5}{ppm} and low jitter.
A clock buffer (CDCE18005) distributes the clock to all devices, isolating the lines.
For the frequency exploration, we define: \(f_{\text{min}} = \qty{1}{\mega\hertz}\), \(f_{\text{max}} = \qty{200}{\mega\hertz}\), \(\Delta f_{\text{b}} = \qty{10}{\kilo\hertz}\), and \num{500} payload iterations.

We evaluate the devices under buffered and unbuffered configurations.
The former reduces flash stress via a pre-fetch buffer and access wait states, while the latter disables them, helping isolate the limiting subsystem.

\subsection{Degradation of Maximum Error-Free Frequency}

The \gls{mef} of a device depends on the propagation delay and critical paths that the payload activates.
An ideal payload fails at the lowest possible frequency by uncovering the most critical paths, making \gls{mef} the most important feature in the comparison framework.
\autoref{fig:degradation_err_free_freq_vs_temp} shows in orange the median \gls{mef} across devices with a quasi-linear relation to temperature.
Although the degradation in \gls{mef} at elevated temperatures appears across configurations and payloads, it is more pronounced in the unbuffered configuration.

The CPU test reaches the highest frequencies, which we attribute to minimal memory access, making it the most resilient and least informative on degradation.
Conversely, the matrix payload has better level of critical path activation with errors at lower frequencies, particularly in the buffered case.
This is likely due to the cross-subsystem nature of the test, involving flash access, RAM access to hold data, and CPU usage for multiplying.
In the unbuffered configuration, the RAM R+W and flash tests have errors at the lowest frequencies, attributable to heavy flash access for code.

To compare configurations independently of absolute frequencies, \gls{mef} degradation in \autoref{fig:degradation_err_free_freq_vs_temp} at temperature \(T_i\) is calculated with respect to \(T_{i-1}\) and relative to \gls{mef} at \(T_0\) (\qty{20}{\degreeCelsius})
\begin{equation}
    D\left(T_i\right) = \frac{\text{MEF}\left(T_{i-1}\right) - \text{MEF}\left(T_i\right)}{\text{MEF}\left(T_0\right)} \cdot 100
\end{equation}

All payloads show degradations of around \qty{2}{\percent} per \qty{10}{\degreeCelsius}-step.
Most tests have a decreasing variation of frequency with temperature, possibly due to the inverse dependence on mobility.
Non-linearity is stronger on the unbuffered setting, where step-wise degradations vary more.
In general, CPU and RAM march tests are less reliable with larger spreads.

When unbuffered, flash and RAM R+W tests have the highest degradation (\qty{13.7}{\percent}), and a steady evolution.
Conversely, the CPU test shows a degradation of \qty{11.8}{\percent} concentrated in the initial steps.
Degradation while buffered is more uniform, with flash and matrix tests showing the smallest spreads and steady \gls{mef} reduction.
While the CPU test is the most degraded with \qty{10.7}{\percent}, it is the least affected by the configuration change, only \qty{1}{\percent} compared to the near \qty{4}{\percent} of other payloads.

Results indicate that the most effective probe for detecting \gls{mef} reduction depends on the configuration.
In the unbuffered scenario, payloads heavily using memory show errors at lower frequencies, indicating that flash access is the main timing limitation.
This is likely due to the increased sensitivity to flash latency in this configuration.
In contrast, the buffered configuration reduces stress on the flash, shifting the performance-limiting paths.
In this case, the matrix payload fails at lower frequencies, suggesting that the non-sequential memory access due to execution loops and the intense ALU usage stress the pipeline more.
Considering the evolution of \gls{mef} degradation, both matrix and flash payloads show good properties for monitoring: low spread and \gls{mef} across configurations.

\begin{figure}[t]
    \centering
    \includegraphics[width=\linewidth]{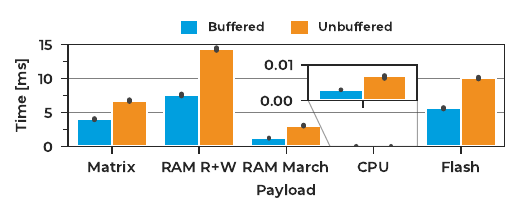}
    \vspace*{-0.7cm}
    \caption{Median execution time of the payloads at \qty{20}{\degreeCelsius}.}
    \label{fig:execution_time}
    \vspace*{-0.5cm}
\end{figure}

\subsection{Error Transition}

The evaluated payloads behave differently when approaching frequencies close to the hardware limitations.
Some run error-free up to a frequency at which execution cannot continue: the \gls{mof}.
Other payloads detect erroneous executions and still resume normal operation afterwards.
The latter have an error transition region, beginning at \gls{mef}, until all executions fail (\cf \autoref{fig:frequency_exploration}).

Results reveal that the CPU and March-C tests have no error transitions.
\autoref{fig:error_transition} shows errors across frequencies and temperatures for the other payloads.
At \qty{100}{\percent}, all \num{500} test executions failed, solid lines are median values, and shades are the \gls{iqr}.
The RAM R+W and flash payloads have a progressive error count only under the unbuffered configuration, and their spread is relatively small across temperatures.
In addition, there is a steep error increase after the \gls{mef}, and a clear separation of curves for temperature steps.
In contrast, the larger spread of the matrix payload on both configurations indicates \gls{mef} variation among devices.

Early signs of erroneous execution are good for a probe payload: when \gls{mef} coincides with \gls{mof}, the search comprises many executions where \gls{dut} hangs, requiring a system reinitialization.
Payloads with a smoother transition between \gls{mef} and \gls{mof} are preferable, as errors indicate a hardware limitation without a reboot.
In this regard, the matrix payload has the best behaviour with transitions on both configurations.

\subsection{Payload Execution Time}

Execution time is the main overhead factor of payloads: shorter tests reduce downtime, enable more frequent testing \cite{2005_Paschalis_EffectiveSoftwareBased}, and allow more iterations, improving reliability.
\autoref{fig:execution_time} shows that median execution times at \qty{20}{\degreeCelsius} are relatively consistent across configuration, differing in a scaling factor.
The RAM R+W test has the highest execution time due to writing, reading, and hashing, while the CPU test is the fastest (under \qty{10}{\micro\second}), benefiting from running directly from RAM with register-only interactions.
Flash and matrix tests fall in between, with the matrix executing \qty{30}{\percent} faster than the flash.

\begin{table}
    \caption{Qualitative evaluation of payloads between zero (poor) and three (excellent) stars according to performance.}
    \label{tab:payload_comparison}
    \centering
    \begin{tabular}{lS[table-format=2.2]S[table-format=2.2]S[table-format=2.2]}
        \toprule
        \textbf{Payload} & \textbf{\gls{mef}} & \textbf{Execution time} & \textbf{Error Transition} \\
        \midrule
        \textbf{Matrix}     & \threestars & \twostars & \threestars \\
        Flash      & \threestars & \onehalfstars & \onehalfstars \\
        RAM R+W    & \twostars & \onestar & \onehalfstars \\
        RAM March  & \onestar & \twohalfstars & \zerostars \\
        CPU        & \onestar & \threestars & \zerostars \\
        \bottomrule
    \end{tabular}
    \vspace*{-0.6cm}
\end{table}

\subsection{Payload Comparison}
\autoref{tab:payload_comparison} qualitatively compares payloads over the discussed features.
\Gls{mef} reflects how effectively a test excites critical paths, with lower frequencies indicating more sensitivity.
Execution time considers how long it takes to complete the test, shorter times reduce overall duration.
Error transition indicates whether early payload failures facilitate the search for the \gls{mef}.
Across all categories, the matrix payload performs well, with low \gls{mef}, short execution time, and early failing behaviour, making it a good option for degradation evaluation.

 \section{Conclusions and Future Work}\label{sec:conclusion}

We presented a deployable online technique for monitoring ageing on \gls{cots} \glspl{mcu} utilizing timing windows to determine the limit of operational frequency.
By exposing devices to degraded conditions, we validated it can detect hardware degradation.
Our evaluation framework allowed comparing test payloads and identifying the best-performing one.
Payload behaviour under multiple configurations provided insights into the degradation effects on different subsystems of the devices.

In future work, we envision extending tests to various aged devices and exploring the automatic generation of payloads.

\vspace*{-0.2cm}

\bibliographystyle{IEEEtran}

\balance

\label{lastpage}

\end{document}